# Enabling Trusted App Development @ The Edge

**Tom Lodge**
University of Nottingham
thomas.lodge@nottingham.ac.uk

**Anthony Brown**
University of Nottingham
anthony.brown@nottingham.ac.uk

**Andy Crabtree**
University of Nottingham
andy.crabtree@nottingham.ac.uk

## ABSTRACT
We present the Databox application development environment or SDK as a means of enabling trusted IoT app development at the network edge. The Databox platform is a dedicated domestic platform that stores IoT, mobile and cloud data and executes local data processing by third party apps to provide end-user control over data flow and enable data minimisation. Key challenges for building apps in edge environments concern *(i)* the complexity of IoT devices and user requirements, and *(ii)* supporting privacy preserving features that meet new data protection regulations. We show how the Databox SDK can ease the burden of regulatory compliance and be used to sensitize developers to privacy related issues in the very course of building apps. We present feedback on the SDK's exposure to over 3000 people across a range of developer and industry events.

## ACM Classification Keywords
H.5.m. Information interfaces and presentation (e.g., HCI): Miscellaneous;

## INTRODUCTION
The predominant paradigm for computing is centred in the cloud. However, as the Internet of Things (IoT) emerges, the requirement to push increasing volumes of data to the network for centralized storage and processing will impact system resilience, network traffic, latency and privacy. An alternative approach is to "extend the cloud to where things are" [14] and shift data storage and processing to the edge of the network. In this model, nodes at the edge perform the bulk of storage and processing, keeping data off the core network, reducing latency and improving the potential for data privacy. The model has gained significant traction over recent years, and the IDC [21] predicts investment in edge infrastructure will reach up to 18% of total spend by 2020.

The domestic space is seeing a growth in dedicated hardware that brings more data storage and processing to the edge [1,4,16,17,31]. Many of these products unify access to connected home devices and provide facilities (voice, web UIs, apps) for automation and control. With new data processing regulation (GDPR) looming in Europe [50], accompanied by growing concern amongst ordinary people about the (ab)use of personal data, we anticipate this space will grow to include new domestic platforms that take a more principled approach to exploiting personal data generated by IoT devices, mobile and cloud services. The Databox platform [51] provides one instantiation of a domestic 'privacy-preserving' edge-based solution, running data processors (apps) within a sandboxed environment where access to and use of data is constrained by 'SLAs' or user-negotiated contracts.

The distinguishing feature of such platforms is that processing moves to the data, rather than data to the processing, and data distribution is limited to the results of local queries enabling the 'data minimisation' that is required under GDPR. Developing apps that run on these platforms is challenging. There are challenges that are already familiar to IoT developers: *(i)* processing data from an increasingly heterogeneous range of data sources, *(ii)* across wide variability in domestic environments each with a plethora of sensors, devices and actuators, and *(iii)* competing systems with inconsistent patterns of behaviour [75], plus *(iv)* the need to support multiple users with diverse requirements.

Add to this the challenge of meeting future data protection regulation and (thereby) gaining user trust. This requires that developers *demonstrably* respond to the requirements of data protection regulation *in* the apps they produce [19]. Furthermore, it is clearly the case that 'developers' is a broad category including makers, hobbyists, and enthusiasts. Development environments are therefore needed that enable data protection across a broad cohort while providing, as Newman [67] argues, developers and end-users alike with the tools they need to build the (often niche) functionality that they require.

This paper has two main contributions: *(i)* the design, implementation and evaluation of an edge-based application development environment enabling a broad spectrum of developers to address the challenge of building apps for heterogeneous domestic environments, *(ii)* in ways that enable compliance with key features of new data protection regulation to engender end-user trust.

## RELATED WORK
Development of the Databox SDK is motivated by 3 interconnected areas of work: *(i)* Domestic smart hubs, *(ii)*

privacy, data utility and trust, and *(iii)* developer support and end-user programming.

**Domestic Smart Hubs**
The multitude of different standards, network and data protocols employed within the domestic IoT space has resulted in the emergence of IoT ecosystems aimed at providing *(i)* interoperability across devices *(ii)* control interfaces for device management, and *(iii)* support for home automation. Within the open source community, many IoT systems have also been designed to run on local hardware, whether ARM, x86 or embedded system such as Arduino and Raspberry Pi [11,15,17,18,24,34,35,36]. These systems are aimed at technically competent users and are underpinned by programming frameworks to support further extension.

There is also a highly competitive startup scene, with a range of products on the market aimed at the general consumer [7,12,31,46], typically offering easy integration with IoT devices and polished control interfaces. The most significant inroads have been made by the large Internet companies. Amazon's 'Echo' [1] is installed in tens of millions of households, for example, and Google's 'Home' [16] is gaining market share as is Apple's HomePod [4]. These systems perform some local storage and processing as a means of reducing latency and reliance on an upstream network, but still use companion cloud-based systems when needed. However, the mechanisms and processes utilised by these cloud systems remain opaque to the end user. Not only is there a *lack of transparency* around the flow and use of data, there are a notably few tools for end-users to *restrict* data flow or exploit it for individual purposes.

**Privacy, data utility and trust**
One common approach to enabling transparency and end-user control is to exploit an 'infomediary' to *bridge* between third parties and user data. Personal Data Management Services, whether cloud-based [30] or at the edge [20], store consumer data and provide explicit contracts to underpin data exchange. Nonetheless, critics have suggested that data privacy is a 'secondary feature' for motivating the uptake of such systems [66,74], and that *utility* (i.e., enabling users to exploit their personal data to their own ends) is the primary driver (hence our app-driven approach). Responding to the *social context of use* is also critical to building trust in the personal data ecosystem [66].

Take, for example, the seemingly straightforward action of installing an app that will make use of personal data. The legal requirement that 'notice and choice' or informed consent be enabled at install time turns upon the end-user having a clear and meaningful understanding of the data, including what it is, who wants it, what for, where it will be stored, who else will have sight of it, what it reveals, what risks attach to it, etc. Much of this 'meta information' is mandated by new data protection regulation. However, it is bound up in an app's algorithms and supporting systems and must, therefore, be made available by *the app developer*.

As one popular online web developer magazine [19] puts it, GDPR will expect developers to be more transparent about the ways they collect and use data, more considerate of their users, and more thorough in their development and documentation processes; they will be expected to adopt the principle of data minimization and ensure that users are informed about the data flows containing their information, and their rights over them, and in the most privacy-positive ways possible. New data protection regulation is putting the job of building end-user trust on to the shoulders of developers. *How* is compliant app development to be supported and data protection principles to be built-in to the smart home development ecosystem? *What* tools and services should be built to help developers get to grips with the demands of the contemporary social context in which they operate?

**Developer support and end-user programming**
The matter of developer support is not straightforward. Commercial and open source ecosystems provide development environments that support the creation of new product integrations or bespoke functionality oriented around a product's features [2,3,17,24,34,39], and are typically targeted at competent and/or professional programmers. However, Newman [67] has noted that the burgeoning array of connected domestic devices makes it intractable for developers to build applications to keep pace with the needs of users. He thus argues for the need to support end-user programming to allow a diverse cohort of people to "compose the functionality that they need". Others strengthen this viewpoint, pointing to a mismatch between the intentions of vendors and the expectations of users arising from varied contexts of deployment [60].

The academic literature has a long history of research into systems promoting end-user agency in domestic environments [54,55,56,63,68,71]. End-user programing has been employed very successfully to connect IoT and web services together for task-automation [65], and many commercial services [22,23,40,48,49] now offer some variant of the *trigger action* paradigm [56,68]. The most popular, IFTTT enjoys a considerable user base [65,72]. All of these systems employ graphical environments, allowing non-coders to set up custom logic ('applets', 'zaps', 'routines', 'scenarios' or 'flows') to perform simple automation. Some commercial hardware offerings harness similar environments for end-user coding [46,31,45,47]. These services are subject to the ever-familiar tradeoff of *simplicity vs. expressiveness*. Trigger-action programming is also unidirectional, making it difficult, for example, to further react to the result of an action, and a range of other problems attach to the approach [59,60,62,65,72].

Trigger-action programming is a variant of the more *flexible* Flow Based programming (FBP). FBP models a system as a directed graph of black boxes with well-defined input and output interfaces (ports) connected by edges. FBP is principally concerned with information flow: data (information packets) flows along edges, in and out of

nodes, and is processed and modified along the way. FBP has several useful features:

- Data flow is separate from program logic, making it simpler to understand, unencumbered by the details of node processing.
- Nodes are re-useable and can be endlessly reconfigured.
- Flow based programs can use different languages for the internal implementation of node logic and the connectivity between them.
- Programs are generally easy to reason about.
- Interfaces for construction of flow-based programs can be very simple.

FBP inspired environments have surfaced in a number of domains including general programming [32], web development [13], embedded systems [27,33], and machine learning [6,41]. One popular example of an FBP-inspired environment is IBM's node-RED, an open source toolkit for developing code for embedded and IoT systems, which has also been repurposed in other domains [57,61,69]. node-RED is a single user environment that consists of two parts: *(i)* a visual web-based programming interface, and *(ii)* a lightweight, single threaded execution environment. With node-RED, a node will typically generate new data (an *input* node), or process data (a *processing* node) or perform an action (an *output* node). A flow typically, but not necessarily, consists of one or more of each of these node types. Data typically passes through a flow from left-to-right, though there is no requirement for it to be unidirectional.

**Summarising Key Challenges**

Development of the Databox SDK is motivated *a)* by the need to make personal data flows transparent and amenable to end-user control; *b)* by the need to enable privacy and data utility and to help developers construct apps that comply with key requirements of new data protection regulation; and *c)* by the need to support a diverse constituency of developers, including potential end-users themselves. These motivating factors provide 4 key requirements for the SDK, described in Table 1.

| REQUIREMENT | description |
| --- | --- |
| Inclusivity | The SDK should support a diverse cohort of developers |
| Data utility | The SDK should allow developers to exploit heterogeneous sources of data |
| Expressiveness | The SDK should enable non-expert programmers to create rich functionality |
| Trust | The SDK should help developers build data protection into apps |

**Table 1. Key SDK Requirements**

## THE DATABOX SDK

The SDK is one component of the wider Databox ecology, which consists of an app store, a local execution environment, and web and mobile user control interface for installing and managing apps. Apps built by our SDK can be published to the app store and subsequently installed and run locally 'on the box'. To address the inclusivity, data utility, expressiveness and trust challenges we substantially re-engineered node-RED, including a full rewrite of the front-end editor and new server functionality.

**Inclusivity**

To meet to the requirement that our environment supports a diverse user base, our early goal was to present the underlying components of the Databox system as set of relatively simple abstractions. In addition, as with environments such as IFTTT, we wanted to ensure that users could 'get going' with minimal effort, i.e., without the need install and configure software to build apps or install a Databox to run tests.

*Mapping data stores to processors and outputs*

The SDK abstracts the Databox platform architecture into three 'node' types: *datastores*, *processors* and *outputs* (Fig.1). *Datastores* represent all devices (or services) that create data and are often the first node in a flow; they are analogous to triggers in the trigger-action paradigm. *Processor* nodes are functions that operate on data, and it is here that custom behaviours and logic are encoded. *Processor* nodes typically consume one or more inputs and send results to one or more outputs. *Output* nodes perform an action, such as actuation, visualization, or data export.

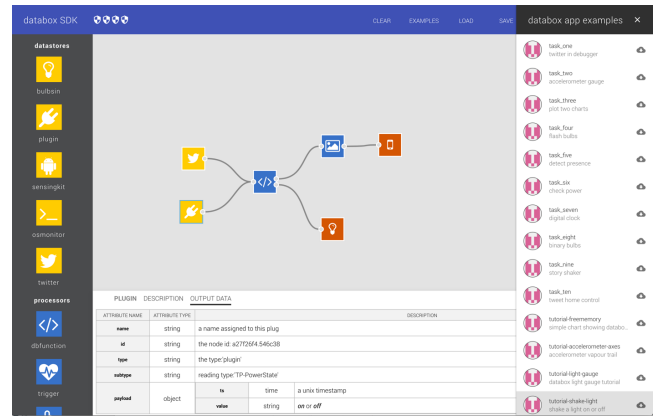

**Figure 1. Databox SDK main screen**

Once deployed, the *datasource* nodes interact with the Databox platform API to request the tokens required to access data according to the terms of a user-configurable Service Level Agreement or SLA. This functionality is transparently provided by the SDK, insulating developers from the detail.

*Testing framework*

node-RED provides a few tools to help debug flows, including a *debug* node, that can be connected to any output to inspect data. However, node-RED is a single user system, designed assuming the development and runtime environment cohabit the same (often embedded) device. This means there is no distinction between 'testing' and 'deployment'. To test that a flow works, it must be deployed and inspected using the logging and debugger tools.

To support a multi-user environment, we needed to (*i*) introduce user accounts, (*ii*) ensure isolation between the flows developed, tested and deployed by each user and (*iii*) support separate 'testing' and 'deployment' workflows. To achieve *(i)* we authenticate users through Github[1] oAuth and request permissions for read and write access to a user's public Github repositories; this has a secondary advantage of allowing us to provide "out-of-the-box" code management (i.e. all of a user's flows correspond to Github repositories within their account). To achieve *(ii)* and *(iii)* we use Docker[2] to 'containerise' separate instances of the node-RED execution environment for each user.

When a user initiates testing, a new instance of the node-RED execution environment is built as a Docker container to run the flow. Each instance can be 'specialized' at build-time to add any dependent libraries.

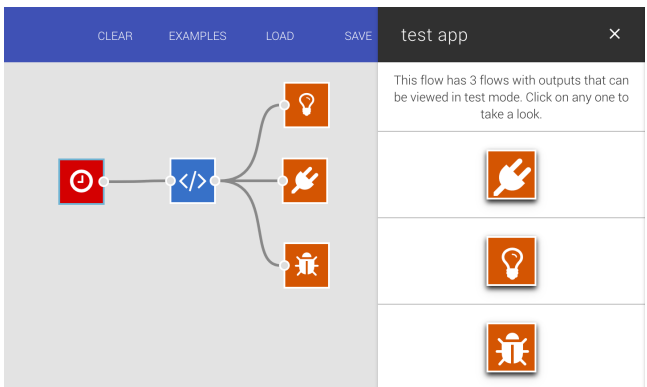

Figure 2. Testing outputs

Once built, the user is presented with a list of outputs that they can inspect (Fig.2) and all datastores generate mock data to trigger changes in the outputs. Once tested, the app can be published to an app store, where it can subsequently be downloaded to a local Databox.

**Data utility**
In the introduction we noted the challenge of processing data from a large and heterogeneous set of data sources. We have built a set of features aimed at helping developers to work with data, i.e., to inspect its structure and attributes under different profiles (context sensitive help), to examine how it changes over time (provision of input and output schemas), to pre-populate functions with skeleton code (translation of schemas to Javascript variables), and provide development-time checking of schema compatibility (static type checking). We will now consider these in more detail.

*Data schemas*
*Datastores* commonly present a set of choices to the developer that will impact upon the emitted data *type* (i.e. the particular schema of the data) and sampling *granularity* (GDPR recommends that data processors offer users *granular choice* over data sampling rates). As data moves through a flow it must mutate to conform to the schema expected by downstream nodes. For example, a node that calculates a linear regression function might expect its input data to consist of a an array of *x,y* tuples whereas a node that turns a bulb on or off may only require a single *true* (on) or *false* (off) input value. As a result, developers must continually ensure that any downstream incompatibilities borne from an upstream configuration change are successfully resolved.

With node-RED, incompatible connections will not be discovered until runtime, i.e., the environment offers no development-time checking. We have extended node-RED to require that nodes formally define the *type* of the data they produce and expect (as a json-schema[3]). This provides several advantages. First, any incompatible connections can be flagged up during development; if an upstream node is re-configured, the environment can flag any resulting downstream incompatibilities. Second, we can tailor the configuration interfaces of downstream nodes to reflect the structure of incoming data (and void any configurations that no longer make sense). Take for example our 'chart' output node that builds real-time data plots. If we have the following flow in Fig.3, then data from a smartphone (the yellow node) flows into a chart (the blue node), and finally out to a display (the orange node). The smartphone offers a variety of sensors, including an accelerometer, magnetometer, battery level indicator, output from Bluetooth scans, microphone level and so on. The data emitted from the smartphone node could therefore be, amongst others, a matrix of floats (e.g. *x, y, z* accelerometer values), a single float within a range (battery level), an array of string tuples (Bluetooth scan) and so on. Because we ensure each smartphone sensor has a datatype schema, the SDK can automatically adapt the chart node's configuration options (i.e. appropriate charts, data to be plotted) to align with the expected input data.

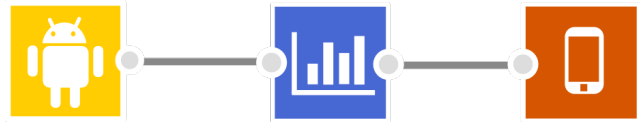

Figure 3. SDK flow

*Support for writing code*
The SDK allows users to write raw (JavaScript) code using our *function* node. Given our schema definitions, the SDK is able to provide additional coding support to users. Without type definitions or knowledge of the input and output nodes, a developer must trace through the program flow to work out the format of the data coming in to the function and the format of the data expected by nodes it outputs to.

---

[1] www.github.com
[2] www.docker.com
[3] http://json-schema.org/

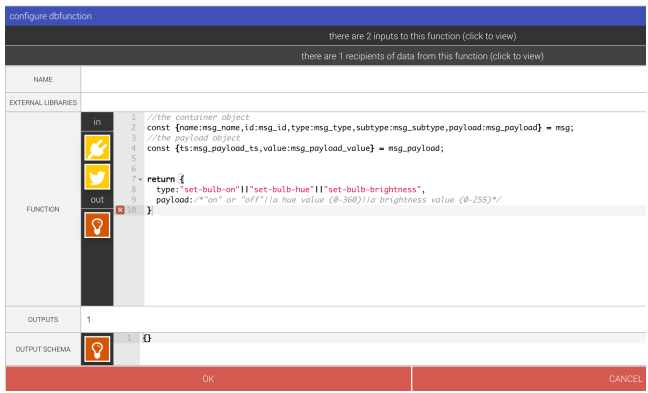

**Figure 4. Support for writing code**

With type definitions we are able to clearly signpost the inputs and expected outputs and even auto-generate skeleton code (Fig.4) to assign variables to input data and create stub outputs that are compatible with the downstream node(s). We are also able to utilize Facebook's static type checker, Flow,[4] to perform dynamic type checking: we convert the json-schemas of connected input and output nodes into a Flow function signature, and use this to re-validate code as it is written by users.

Our environment, therefore distinguishes itself from node-RED by *i)* providing each node with access to the details of all nodes that it is connected to, and *ii)* mandating that each node specifies the type of data that it expects as input and the type of data it outputs *for every possible configuration*.

*Context sensitive help*

| | SENSINGKIT | DESCRIPTION | OUTPUT DATA | PERSONAL DATA |
|---|---|---|---|---|
| light | | | | |
| Measures the ambient light level (illumination) in lux captured by a device camera. The following is an indication of value ranges | | | | |

| lux value | description |
|---|---|
| 0.0001 | Moonless, overcast night sky (starlight) |
| 0.002 | Moonless clear night sky with airglow |
| 0.05 to 0.36 | Full moon on a clear night |
| 3.4 | Dark limit of civil twilight under a clear sky |
| 20-50 | Public areas with dark surroundings |
| 50 | Family living room lights |
| 80 | Office building hallway/toilet lighting |
| 100 | Very dark overcast day |
| 320-500 | Office lighting |
| 400 | Sunrise or sunset on a clear day |
| 1000 | Overcast day |
| 10000 to 25000 | Full daylight |

**Figure 5. Context sensitive help for light sensor**

In the course of developing the SDK we found that the provision of data schemas alone was not sufficient to establish a clear sense of the data we were working with. For example, knowing that a light sensor emits a *timestamp* and *lux* reading in the range of *0* to *130000* provides no indication of what the values correspond to, e.g., does a reading of *20000* mean it is very bright or dark? Similarly, if a developer wishes an app to respond to accelerometer data from vigorous shaking or walking or jumping, what are the expected ranges of the *x, y* and *z* components? There is missing context. All our *datasource* nodes therefore provide further contextual help that elaborates, where appropriate, upon the data schemas. The information includes, for example, the rate(s) that data may be generated at and the expected range of values under different contexts (Fig.5).

**Expressiveness**

Though FBP provides for greater expressiveness than the simpler end-user environments we considered earlier, more effort is required to ensure that our environment can support rich functionality *outside of code*. Richer functionality is a feature of *(i)* the variety of nodes provided, and *(ii)* the richness and flexibility of node configuration options. With regard to *(ii)*, take for example, apps that require some element of data visualization [60]. A familiar set of tools for working with graphics might include amongst others: a canvas, layers and shape manipulation tools (rotate, scale, colour, etc.); i.e., a fully featured, domain-specific development environment in its own right. To support this richness, we ported the current, jQuery [25] based front-end code (used for building simple form-based configuration interfaces in node-RED) to a declarative framework (React [37], Redux [38]) more suitable for building complex interfaces.

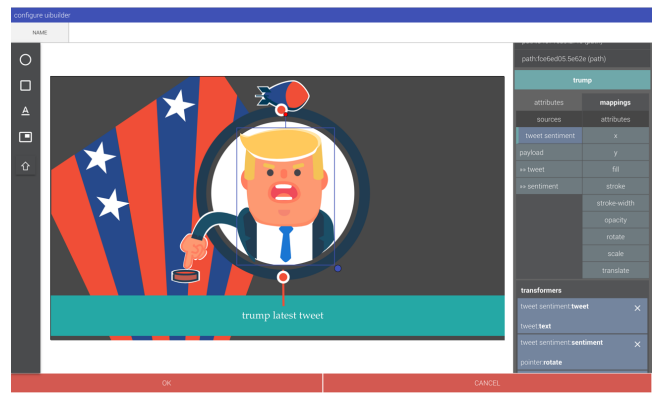

**Figure 6. UIBuilder interface**

By way of example, consider our visualization node (*UIBuilder),* which enables developers to rapidly construct high quality, data-driven infographics. The node provides a palette and canvas for constructing, manipulating and importing scalable vector graphics (SVG). The configuration interface presents the attributes of all of the data from connected input nodes and allows users to 'map' any item of data to any of the attributes and styles of the components from which an SVG is composed (colour, size, position, stroke, etc.). In a simple example, the lux value of a light reading might be attached to the radius or transparency of a circle. Fig.6 depicts a more complex example where the rotation of Trump's arm is connected to the sentiment score of data from his latest tweet. *UIBuilder* also provides a set of advanced options for mapping functions to data transforms and cloning or removing objects under particular conditions. Note that our framework for building new nodes for the SDK is aimed at developers, and is not expected to be used nor required by users of the SDK.

---
[4] https://flow.org

**Trust**

The features thus far described have principally been concerned with support for building app functionality. In this section we detail a novel set of features introduced to help developers address the concerns of GDPR and build user trust by *(i)* creating GDPR compliant contracts *(ii)* tracking personal data, *(iii)* providing facilities for runtime inspection, *(iv)* and exposing risk.

*Creating GDPR compliant contracts*

When a user seeks to install an app on the Databox they are presented with a Service Level Agreement (SLA). This consists of a multi-layered notice that furnishes the information required to be provided to 'data subjects' under GDPR (articles 12-18) in an easily readable format (see [53] for further details). SLAs also enable end-users to exercise granular choice over data sampling and reporting frequency, where applicable. SLAs are not static notices then, but dynamic, user-configurable consent mechanisms that surface and articulate who wants to access which connected devices and what they want to process personal data for. They are constructed from a *manifest* file, which is submitted alongside an app when it is published to the app store. The SDK streamlines this process; given its knowledge of an app's construction it already knows the data sources being accessed (and at which granularity) and the outputs, all of which can automatically embedded in the manifest. All that remains is for the developer to provide a description of the app and its benefits, and to provide the statutory information required by GDPR.

*Personal data tracking*

In addition to requiring nodes to provide a schema defining input and output types and data descriptions, we also require that each node provides a schema detailing the *personal data* that it outputs (Fig.7). Inspired by GDPR, our schema distinguishes between three classes of personal data: *(i)* 'identifiers' such as internet protocol addresses, cookies, radio frequency tags, etc., *(ii)* 'sensitive' data or data pertaining to sexuality, biometrics, health, genetics, race, etc., and *(iii)* and data that is clearly 'personal' including finances, age, gender, interests, location, vocation, employment status, behaviour and consumption patterns, etc.

Within each of these classes of data, we also distinguish between primary data, i.e., data which is evidently personal, and secondary data which enables personal 'inferences' to be made. One challenge we anticipate developers will face is how to account for the subtler inferences that may be performed on data [42] or combinations of data (sexuality, age, gender, emotional state, etc.); examples include processing of accelerometer to infer gait, or Bluetooth scans to reveal social connections or routines. All 'secondary' (i.e., inferred) data in our personal data schema includes a 'conditions' tag for defining the circumstances under which particular inferences are possible (for example, threshold data granularity, or additionally required data).

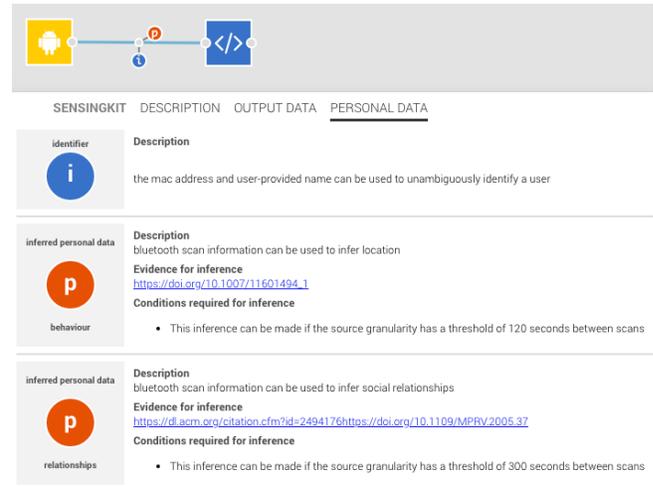

**Figure 7: Personal data descriptions**

Using this schema, we are able to help developers track personal data as it flows through an app. This is a powerful aid to enabling privacy-preserving app development as it makes personal data processing *explicit*. One valuable emerging feature is immediate feedback on the effects of changes in the configuration of upstream nodes upon downstream nodes (e.g., adjusting sampling granularity of a source, or deleting a link).

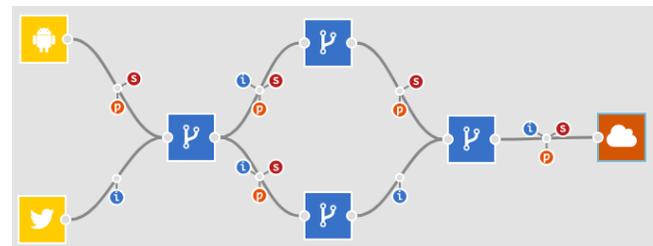

**Figure 8a. Tracking personal data flow (i)**

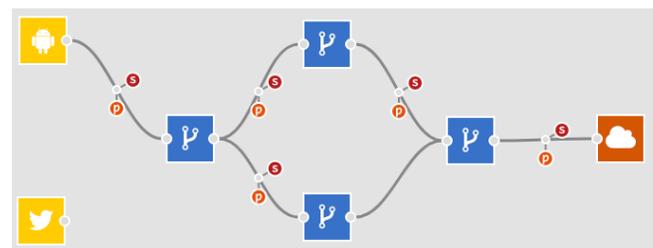

**Figure 8b. Tracking personal data flow (ii)**

Consider Fig.8a, an illustrative flow where data passes through a set of blue *extract* nodes which combine or remove attributes of the data flowing through them.

*Personal*, *sensitive* and *identifier* data circles (the orange *P*, red *S* and blue *I* circles) are used to label the edges between nodes. Contrast this with Fig.8b, which shows how, when the edge between the lower (yellow) twitter node is removed, the edges of all implicated downstream nodes are automatically adjusted to indicate that no personal *identifier* data is now being used.

*Runtime inspection*

Though the development environment ensures that the sources of data that an app operates on and what it outputs to are made transparent, the way in which the app operates, i.e., how a decision is arrived at, or how a data flows through an app remains opaque to a user who has installed the app. This becomes an important matter to surface under article 13 GDPR, which mandates that meaningful information about the logic involved in automated processing be provided to the data subject. To deal with this, apps built in the SDK record the path and state of all data as it moves through a flow. SDK apps are bundled with an interface that uses this path information to make apps 'inspectable' at runtime, thereby surfacing the *provenance* of a particular decision/action.

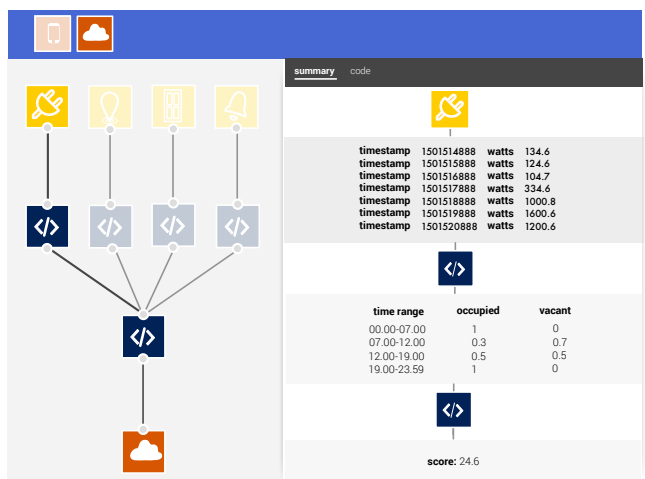

Figure 9. App inspection interface

By way of example, Fig.9 shows the inspection interface on the Databox for an app that processes a variety of IoT data. At the top of the interface are all outputs associated with the app, in this case, a display and data export. Clicking on each output reveals the path taken by data before it reaches the output. A user can select any node in the path to get a *real-time* feed of the data entering and exiting a node, though they may also scrub back and forth to examine historical data. This is a nascent first step towards satisfying the requirements of article 13. We find similar approaches elsewhere [e.g.,73], and others are exploring novel 'comic strip' visualisation techniques to more effectively communicate the logic of automated processing to end users [e.g.,70]; SDK-built apps provide enough runtime information to support development of new visualizations such as these.

*Exposing risk*

A core requirement of any privacy-aware platform is provision of clear guidance on the *privacy risks* associated with any app. Privacy risk relates to the degree of sensitivity and exposure of data used by an app. With IoT platforms there are also additional *operating risks*, relating to the risks of actuating (principally physical) devices (e.g., opening and closing a garage door in response to some trigger). These latter risks are beyond the scope of GDPR but clearly must be identified to users and developers. Identifying risk is challenging, given that it can be introduced by any individual component of a system (i.e., hardware, such as sensors / actuators and software such as apps and drivers) as well as arbitrary combinations of hardware and software in a particular operating context.

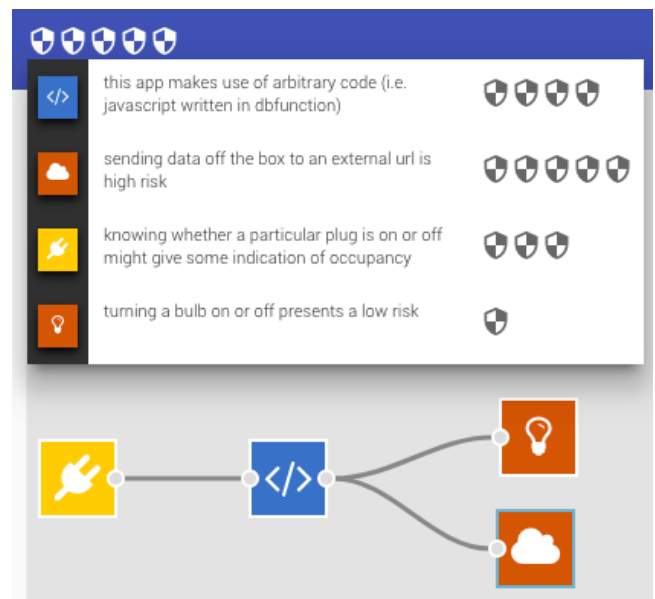

Figure 10. SDK risk overview

Though by no means infallible, our development environment generates a risk rating for apps, based on the aggregate risk of the nodes from which it is composed. Each node in the development environment has a pre-defined *spectrum* of risk. The final risk rating assigned to the node will sit within the spectrum and will be determined by how the node is configured (e.g., the hardware it works with, the proposed data rate, the particular actuation to be performed). Those characteristics of a node that will most influence its risk are *i)* whether it exports any data off the box, *ii)* it triggers physical actuation, *iii)* it utilizes insecure / leaky / non-compliant hardware, *iv)* it uses unverified code or libraries. These are initial first steps at articulating risk, and the SDK provides developers with a view on potential risk in the course of app construction (Fig.10). The risk rating of an app is also made available to users on the app store to further motivate and drive the development of low risk and even 'no risk' apps.

**EVALUATION**

The Databox SDK, like the underlying platform itself, is very much a work in flight and not amenable as such to usability evaluation in any meaningful way. Indeed, following Greenberg and Buxton [58], usability evaluation may even be considered harmful at this point time, leading to a focus on the 'local hill climbing' problem and (thus) 'getting the design right' at the expense of innovating and developing a potentially useful rather than usable system. Greenberg and Buxton suggest that the contrast between usability and usefulness begs a key question for developers: how can we create what could become culturally significant systems and interfaces if we demand they be validated *before a culture is formed around them?*

The answer for Greenberg and Buxton is that instead of treating novel systems and interfaces as prototypes by default whose usability might be straightforwardly measured, we treat them as 'sketches' that illustrate the essence of an idea but have rough or underdeveloped aspects to them. Systems created as interactive sketches provide a tangible means of making design ideas concrete *and* enabling collaborative reflection, discovery, and refinement.

> " … sketching is about 'Getting the Right Design'. Only afterwards does one work on 'Getting the Design Right' … Thus sketching is akin to a heuristic that helps one move closer to the global maxima by circumventing the local hill climbing problem … … … [sketches] should be judged by the questions being asked, the type of system or situation that is being described, and whether the method the inventors used to argue their points are reasonable."

If evaluation is not a matter for usability experts, to whom might we turn to validate our work then? We have no global panacea to offer, no silver bullet as it were, but have sought in our own case to reach out to variety of stakeholders to explore the potential usefulness of the SDK and how it sits within the Databox ecology to enable privacy-preserving application development. In doing so we have sought not only to elicit feedback and constructive criticism regarding the potential usefulness of the SDK, but to form a culture around it in order to foster broad interest and awareness and drive appropriation. Thus our method of evaluation consists in *event-based engagement with a diverse constituency of stakeholders*, including in this case developers, privacy champions, and industry leaders.

**Hackathons**

The developer community is an obvious constituency to reach out to and we have sought to engage it through hackathon events. The SDK has been exposed to both advanced and novice developers at dedicated hack days organised by ourselves [9] and at larger events, including the Mozilla Festival or Mozfest [28,29]. We estimate ~200 developers have been directly exposed to the SDK at hackathon events. Unsurprisingly developer feedback tends to focus on technical characteristics of the SDK. Overall, developers find the programming paradigm and concept of linking together nodes to build functionality easy to understand. Indeed, given the prevalence of the FBP metaphor they are often able to draw useful analogies with other visual coding environments as they learn to get to grips with the specific features of the SDK.

This is not to say the SDK is unproblematic. Novice developers, who we see as key users of the SDK, initially struggled with the event-based nature of app development (i.e., that logic within a node is only triggered upon receipt of an item of data), and when asked what they find most challenging about the SDK, the most common answer amongst novices was "JavaScript". Currently, the SDK is still heavily reliant on the catch-all 'function' node to create custom logic and any reliance on this node will tend to move required competencies beyond the scope of a novice end-user. More seasoned developers, on the other hand, wanted a clearer "at a glance" picture of data flow; they felt that too much detail was hidden behind each node and that a top-level representation of input and output data attributes would be beneficial.

Through hands on engagement with expert and novice developers we are learning what works for 'developers' in the round and how to improve the SDK, though one currently insurmountable issue is of note. It is not to do with the design or functional characteristics of the SDK, but what one might *do* with it. As one developer summed it up, "coming up with an idea of what to build is half the trouble." In more formal terms, it might be said that the SDK rubs up against the 'selection barrier' [62]. Just as mobile phones were once new and app development a novel concept, then so it is with home hubs and personal data stores; evidently *developers* will need time to acclimatise to the opportunities enabled by new approaches that open up application development in heterogeneous environments.

**The Mozilla Festival**

Mozfest attracts a diverse cohort of participants, not only developers but researchers (academic and industrial) from a wide range of backgrounds (ethics, law, art, design, etc.) who have an interest in championing a healthy digital ecosystem that benefits all. Our engagement activities have not been restricted to hackathons then, but also reach out to the broader community. We have thus engaged in public understanding of science activities to develop awareness and promote discussion and debate (Fig.11).

Exploiting open events such as Mozfest has also given us the opportunity to create engaging experiences with our industry partners to demonstrate the Databox ecology and innovative data-driven applications. The Kitchen Databox Demo [43,44] produced with our partners the BBC engaged Mozfest attendees in a novel recipe-based experience where data from Internet-enabled utensils and kitchen appliances drove the delivery of media content providing instructions for making chocolate pots. Not only did the participants get to experience (rather than just hear about) novel data-driven applications, they also got to eat the results!

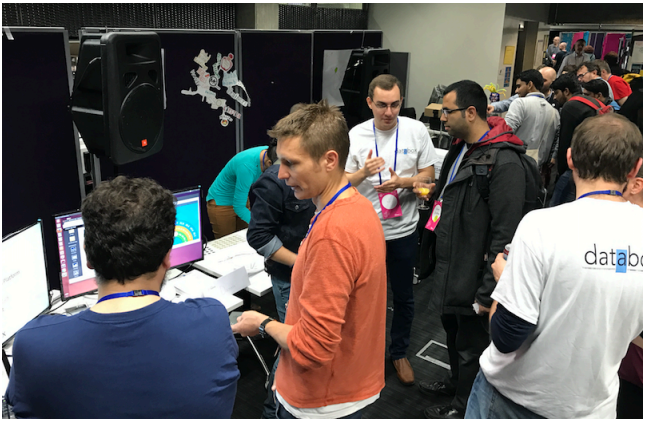

**Figure 11. Community engagement at Mozfest Science Fair**

We estimate ~1000 people have been exposed to the Databox ecology and SDK through outreach and experience-based events, mainly at Mozfest but also at public sessions held at the Annual Databox Symposium. Given the nature of these events, feedback has focused less on the mechanics of the SDK. This is not say the SDK was not discussed, only that the discussion centred on broader matters, such as the perceived relevance of a visual coding approach to enabling a diverse constituency of developers. As one participant put it:

> "I think it's great that there are people looking at alternative paradigms of programming. There's no reason to be complacent when the current method is such a bottleneck for people."

Feedback in this context tended to focus on 'bigger' issues, rather than the nitty-gritty of SDK functionality and interface design. *Consent* was of notable concern to participants, who found traditional models lacking and the SLA enabled by the SDK a plausible and realistic alternative responding to key requirements of GDPR, including legible information and granular choice allowing end-users to actually exercise control. *Risk* was another large-scale concern that was frequently raised by participants, who saw the SDK as taking positive steps in flagging risks to developers and well as users.

**Industry Events**
A third strand of our evaluation strategy has been to engage industry. Building on and exploiting our research links, we have been able to attract several large industry partners including the BBC and BT, two of the largest and most recognised brands in the UK. We have attracted other large industry partners too, but disclosure agreements constrain what we can say and point to the challenges involved in engaging industry players. Nonetheless, it can be a very effective means of exploring the potential usefulness of *and* forming a culture around new design ideas to foster appropriation.

Not only have our industry partners enabled awareness-raising activities (demonstrations and discussions) at high profile internal events, they have also helped us reach out to new constituencies. Our partnership with BT has, for example, not only allowed us to understand their orientation to the Databox ecology and SDK, which has been very positive and indeed formative, but also enabled us to engage a much broader constituency through the company's Innovation Showcase [5], which attracts ~5000 participants from a wide range of commercial sectors and key decision-makers in industry and government.

We do not claim that the SDK has been exposed to all Showcase attendees, but as part of the 'Smart World' showcase, which ran for a full 5 days, we estimate ~2000 people from a wide range of commercial backgrounds rubbed up against the SDK. We seeded the event with an app that visualised sentiment analysis on real time tweets. Individual elements of the visualisation could be clicked on to view how the twitter data was processed and how it influenced the animation, and was used to articulate the SDK and the possibilities it enables.

The SDK was broadly and positively understood by participants, though much like Mozfest, feedback tended to centre on 'big' issues, with GDPR looming on the horizon being by far the most prevalent. A great many participants saw the SDK as a potential means of building *transparency* into the processing of personal data and of thereby meeting key requirements of GDPR. We also found that, unlike developers, industry participants could readily imagine *application scenarios* spanning insurance, banking, advertising, medical trials and managed housing.

**Building a Culture of Appropriation**
Our efforts to engage stakeholders and build a culture of appropriation around the SDK and Databox ecology in which it sits have taken place over a 2-year period, and are ongoing. They may seem trite to a usability engineer armed with a battery of measurements in place of a handful of anecdotes. Nonetheless, these activities have had a tangible impact on the development of the SDK and driven iteration. For example, our visualization node and use of data schemas was developed in response to feedback to our first Mozfest hackathon. Indeed implementation of skeleton code generation, live logging of container builds (during testing) and our *extract* and *trigger* nodes were all inspired by the hackathons. Our risk overview and provenance interface were improved in response to participant feedback at our second Mozfest, and more recent development of the personal data tracking features emerged from feedback from the BT Showcase.

The effort invested in engaging relevant stakeholders appears to us at least to exceed that normally involved in usability evaluation. Not only have engagement activities taken place over a substantial period of time, but considerable effort is involved in their organisation, performance and accomplishment, with each event being expressly designed for its participants, including use cases, demonstrator apps, videos and publicity materials, tutorials and other documentation as needed.

More broadly our approach may be seen as a form of 'in the wild' evaluation [52], which extends beyond the traditional concern (in HCI at least) with 'users' to include other salient parties and expertise. Thus, in this case, our evaluation extends beyond developers (novice, intermediate or expert) to the broader development community and industry players who are capable of *driving change*. For example, that the SDK has been included in Mozfest is due to it engaging with contemporary themes of broad concern, e.g., widespread data harvesting and decentering personal data processing [8,10], and that it has been included in BT's Innovation Showcase is due to a set of concerns implicated in personal data processing that the company wishes to champion to its customers, which the SDK demonstrably addresses.

Our method of evaluation recognises that it is not only, or even primarily (?), developers who will drive the uptake of new approaches and tools but a range of other players, some far more powerful than users of any colour, shade or hue. It is for this reason that our approach is intentionally *selective*, not 'simply' being events-based but basing events in relevant contexts with a *range* of relevant players. Thus, and for example, our work extends to collaboration with the BBC and the creation of engaging public experiences [26] to promote broad awareness of the near future possibilities enabled by the SDK and Databox ecology.

Does culture building dispense with users and usability then? We think not. But it does put emphasis on moving beyond the safety of established comfort zones and proprietary claims on expertise (e.g., that this or that technique is the best way to validate novel systems). While it may be unsettling, disruptive even, we think the matter is far from settled and open to future research.

**Future Research for the SDK**
In addition to concrete changes enabled by stakeholder engagement, a number of interesting challenges have emerged which we are also keen to explore in greater detail and which are, we think, of broad relevance.

*Runtime inspection*
Our work on making the operation and intent of apps intelligible to end-users is at an early stage (i.e., our data provenance visualization interface) touches an area of research in which there is growing interest. As increasing numbers of apps do things for us, and to us, users will require assurances that they operate as expected, especially as algorithms become more complex (e.g., through machine learning). Within the SDK our nodes already provide a rich set of metadata relating to the data they create, but there is also considerable scope for attaching data related to *how they operate*. This, we believe, can be used to good effect in both describing (in plain English) how an app works, and to bake in information that can be utilized by visual inspection interfaces to help users understand how and why a particular decision was arrived at.

*Tracking the flow of personal data*
We have provided an early implementation of an approach to track personal data (both direct and inferred) during processing. There are three exciting directions that we wish to explore in this space: (*i*) personal data 'typing', i.e., formal derivation of the state of personal data as it is processed when flowing through a processing pipeline; (*ii*) consideration of how processors of data might use information on its personal characteristics in novel ways, for example, our visualisation node could leverage the personal data schema of input data to construct privacy aware visualisations, with the option to redact on demand; (*iii*) analysis of personal data flows for apps at runtime to assess the possibility of inference attacks, for example by examining the types of personal data that are being independently sent to the same endpoint by different apps.

*Articulating Risk*
Much of our work with surfacing risk in the SDK is a 'placeholder' to demonstrate how it might be surfaced at development time in order to sensitise developers to risk during app creation. In reality, the nature and type of risk that can be exposed will change at several points across an app's lifetime, i.e., during app creation, at install and during execution. We are interested in how we might more formally describe and represent risk such that it is both intelligible and actionable for end users.

**CONCLUSION**
The emergence of the IoT is driving a shift in data storage and processing to the edge of the network to reduce traffic and latency and to improve resilience and the potential for data privacy. This shift brings with it challenges of exploiting data from an increasingly heterogeneous array of data sources and complying with key requirements of new data protection regulation, including data minimisation and building effective 'notice and choice' mechanisms into data processing to engender user trust.

Building on the Databox platform to enable data minimisation at the edge of the network, we have argued that new data protection regulation (GDPR) raises an unmet challenge in supporting IoT app development that requires: (*i*) a broad cohort of developers be provided with clear information on the risks that attach to the use of personal data as they build their apps and (*ii*) that all necessary features and information are embedded in apps in order that (*iii*) end-users are provided with both the information they need to make informed choices about running an app and the facility to examine an app's operation. Evaluation of the Databox SDK has moved beyond a traditional concern in HCI with usability and instead focused on exploring the potential usefulness of our approach with a range of stakeholders in order to form a culture around it fostering appropriation. The result is that over 3000 people from diverse backgrounds have encountered the SDK over a two-year period, shaping its functionality and motivating continued development.

**ACKNOWLEDGMENTS**
The research on which this paper is based was funded by the Engineering and Physical Sciences Research Council [grants EP/M001636/1, EP/N028260/1].